\documentclass{article}
\usepackage{graphicx} 
\usepackage{subcaption}
\usepackage{booktabs}

\usepackage[colorlinks,bookmarksopen,bookmarksnumbered,citecolor=black,urlcolor=black,linkcolor=black]{hyperref}

\usepackage{geometry}
\title{\fontfamily{ptm}\selectfont \textbf{3D Medical Imaging Segmentation on Non-Contrast CT} }
\author{\fontfamily{ptm}\selectfont \textbf{Canxuan Gang ~~~~ Yuhan Peng}\\
AI Geeks\\
\url{https://aigeeksgroup.github.io}}
\date{}

\begin{document}
\setlength{\parindent}{0pt}
\setlength{\parskip}{10pt}

\maketitle

\begin{abstract}
\begin{center}
This technical report presents a comprehensive analysis of non-contrast CT image segmentation in the context of computer vision. The study revisits a proposed method and explores the background of non-contrast CT imaging and the significance of image segmentation in computer vision. A thorough report of representative methods, including convolutional-based and CNN-Transformer hybrid approaches, is conducted, highlighting their contributions, advantages, and limitations. The nnUNet emerges as the state-of-the-art method across various segmentation tasks. The report discusses the correlation of the proposed method with existing approaches, emphasizing the importance of global context modeling for accurate semantic labeling and mask generation. Future research directions are also discussed, such as addressing the long-tail problem, exploring pre-trained models for medical imaging, and leveraging self-supervised or contrastive pre-training techniques. Overall, this technical report provides valuable insights into non-contrast CT image segmentation and suggests avenues for further advancements in this domain.
\end{center}
\end{abstract}

\section{Introduction}

Medical imaging \cite{hiwase2025can,qi2025projectedex,zhao2024landmark} has long served as a pivotal modality for clinical examination, providing invaluable insights to physicians \cite{1,2}. Among the myriad of imaging modalities available, non-contrast computed tomography (NCCT) stands out as a prominently employed technique owing to its extensive applicability, expeditious acquisition, and non-invasive nature \cite{3,4}. In scenarios where quantitative assessment of lesions or precise radiotherapy planning is required, obtaining pixel-level masks of the target volume assumes paramount importance \cite{5}. These masks enable accurate delineation of regions of interest, facilitating quantitative analysis, treatment planning, and subsequent clinical decision-making processes. Nevertheless, the manual annotation of a substantial volume of CT slices by a human expert, such as a radiologist, is impractical due to the labor-intensive and time-consuming nature of this procedure. Conversely, within the domain of computer vision \cite{6}, image segmentation \cite{7,zhu2025doei,ge2024esa} offers a promising avenue to streamline this endeavor through its inherent capacity for fully automated and systematic simplification. One of the prevailing applications of three-dimensional (3D) segmentation, utilizing voxel representation, resides within the realm of medical imaging. This stems from the fact that medical imaging modalities, including computed tomography (CT), positron emission tomography (PET), and magnetic resonance imaging (MRI), commonly generate volumetric 3D images \cite{8}. Hence, the attainment of precise and accurate 3D segmentation assumes paramount significance in the domain of medical image analysis, encompassing crucial aspects such as accurate diagnosis and effective treatment planning.


In the field of computer vision, convolutional neural networks (CNNs) \cite{9,10,11,12} have emerged as a resounding success in recent decades, progressively supplanting conventional computer vision algorithms \cite{13, 14} and statistical learning methodologies \cite{15}. The automatic learning and feature extraction capabilities exhibited by CNNs through hierarchical layers of convolutional and pooling operations have brought about a paradigm shift in various computer vision tasks, including image classification \cite{9,10,11,12,qi2025medconv}, object detection \cite{16,17,zhao2025peddet}, and image segmentation \cite{18,19,20}. The success of CNNs is further bolstered by the accessibility of expansive datasets, such as ImageNet \cite{21} and COCO \cite{22}, coupled with notable strides in parallel computing and deep learning frameworks. Hence, CNNs have emerged as the predominant approach for numerous computer vision tasks, primarily owing to their inductive biases to capture intricate image features by virtue of translation equivariance, local sparse connections, and weight sharing \cite{23}. The aforementioned properties have also rendered CNNs exceptionally effective in the domain of medical imaging segmentation, where the precise demarcation of anatomical structures and regions of interest assumes paramount importance for facilitating accurate diagnosis and formulating optimal treatment plans. Consequently, the the state-of-the-art (SOTA) models for medical imaging segmentation predominantly leverage CNNs. Prominent examples include U-Net \cite{24} and U-Net++ \cite{25} for the 2D imaging segmentation, and 3D U-Net \cite{26} and nnUNet \cite{27} for the 3D imaging segmentation. These CNN-based architectures have demonstrated superior performance compared to conventional image processing techniques and alternative neural network paradigms within the domain of medical imaging segmentation.

\begin{figure}
    \centering
    \includegraphics[width=0.6\linewidth]{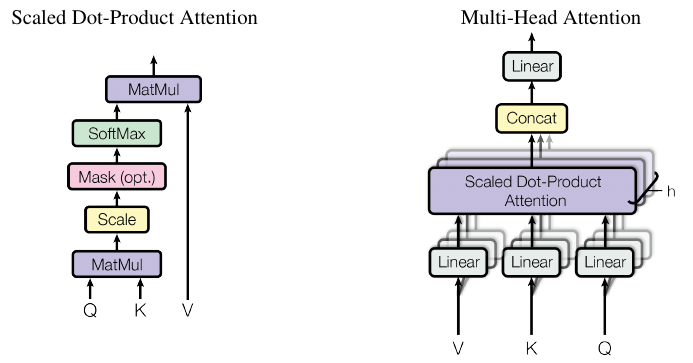} 
    \caption{Self-attention architecture from Transformer \cite{30}}
    \label{fig:1}
\end{figure}

Notwithstanding their achievements in various computer vision tasks, CNNs have exhibited limitations when it comes to precisely segmenting complex and fine structures in medical images. Recent studies have unveiled that CNNs encounter challenges in capturing long-range dependencies and comprehending global contextual information effectively \cite{28}, which are essential for medical imaging segmentation. These limitations frequently manifest in diminished classification accuracy and sub-optimal quality of segmentation outcomes. To overcome these limitations, transformer-based architectures incorporating self-attention mechanisms (illustrated in Figure \ref{fig:1}) have emerged as a promising alternative for various computer vision tasks in general \cite{29}. For instance, the Vision Transformer (ViT) \cite{28} based architecture has exhibited state-of-the-art prowess across diverse computer vision tasks. Notably, it has achieved remarkable performance in image classification \cite{31,zhang2024jointvit,ji2024sine}, object detection \cite{32}, and image segmentation \cite{33}. The recent successful integration of transformer architecture and self-attention mechanisms within the broader domain of computer vision has sparked a burgeoning interest in applying these techniques to medical imaging segmentation \cite{zhang2024segreg,wu2023bhsd,zhang2023thinthick}. Numerous transformer-based models have been proposed in recent research, such as TransUNet \cite{34} for 2D imaging segmentation and nnFormer \cite{35} for 3D imaging segmentation. Despite the promising outcomes exhibited by these models, their reliance on the amalgamation of transformer architecture with convolutional architecture, either for feature extraction or mask decoding, renders them as compromised architectures rather than being entirely "convolutional-free." Consequently, the advancement of a purely convolutional-free architecture built upon transformer and self-attention mechanisms assumes paramount significance. Such developments hold the key to enhancing the performance of semantic segmentation and propelling the field of medical imaging segmentation forward.

Moreover, there are two distinct types of images can be derived from NCCT, namely thick slices and thin slices \cite{36}. Thick slices are acquired through thick scans or retrospective reconstruction methodologies \cite{37}, while the thin slices are acquired from thin scans. Regrettably, publicly accessible datasets containing pixel-wise ground truth are predominantly confined to thickslice datasets, thereby constraining the applicability of existing medical imaging segmentation methods primarily to thick-slice segmentation. Nevertheless, thick slices suffer from inadequate resolution in the depth direction, commonly referred to as z-resolution, resulting in a lack of isomorphism. Conversely, thin slices encompass a greater abundance of volumetric information and maintain isomorphism, thereby bestowing superior segmentation outcomes. Furthermore, solely thin slices possess the capacity to generate authentic 3D volumes through volumetric reconstruction (VR) \cite{38}, whereas the interpolation of thick slices invariably introduces spurious imaging. As aforementioned, the manual annotation of a considerable number of CT slices by an expert human annotator, such as a radiologist, is infeasible given the laborious and time-intensive nature of this process. This challenge is further exacerbated when it comes to annotating thin slices, which involve a larger number of scans. Henceforth, the imperative development of an innovative architecture that integrates effective strategies for segmenting thin slices based on thick-slice annotations assumes paramount significance in the progressive evolution of diagnostic medicine \cite{zhang2024deep}.

In essence, we have introduced a novel pipeline that leverages a fully convolutional-free architecture, employing the self-attention mechanism from the Transformer \cite{30} as an encoder and a lightweight MLP as a decoder. Notably, our pipeline excels in segmenting thin slices by using annotations on thick slices. The contributions of our proposed model can be summarized as follows:

\begin{itemize}
    \item Our research endeavors to establish a novel and proficient architecture that is fully convolutional-free and based on the self-attention mechanism from the Transformer \cite{30} to elevate the the performance of multi-semantic segmentation on 3D medical images, which includes improving the classification accuracy of semantic labels and the quality of segmentation masks. 
    \item Our research focuses on developing a joint loss function that effectively supervises the segmentation of thin slices using thick slices. Specifically, our approach involves utilizing the ground truth of thick slices for direct supervision, ensuring that the predicted mask on thick slices precisely aligns with the ground truth. Moreover, we aim to ensure that the average mask intensity of corresponding thin slices corresponds to the ground truth of thick slices. Additionally, our method seeks to establish a correspondence between the average feature maps derived from corresponding thin slices and the feature map of thick slices.
    \item We aim to contribute to the academic community by releasing a benchmark dataset for thin slice multi-semantic segmentation. This significant undertaking involves curating an extensive collection of NCCT images with thin slices, complemented by meticulous multi-semantic annotations. By making this dataset publicly accessible, we address the prevailing dearth of comprehensive datasets in the realm of medical imaging segmentation. This initiative serves to facilitate researchers in evaluating and refining their models within this domain.
\end{itemize}

This paper endeavors to provide a comprehensive examination and consolidation of the current approaches employed in 3D medical imaging segmentation. Our analysis encompasses a detailed exploration of the respective merits and drawbacks of these methods, along with their significant contributions to the field. By elucidating these aspects, we aim to equip readers with a comprehensive understanding and holistic perspective of this crucial task. Furthermore, we emphasize the significance of positioning our proposed method within the context of prior works, highlighting its unique features and potential advancements over existing methodologies. 
The paper will be organized into discrete sections to ensure a cohesive dissemination of information. The delineated sections encompass:

\begin{itemize}
    \item \textbf{Background}: This section will provide a comprehensive overview of essential background knowledge pertaining to the mechanism of NCCT, as well as the fundamental concepts associated with image segmentation in computer vision. Additionally, it will delve into the examination of recognized benchmarks and datasets that have gained prominence within the field.
    \item \textbf{3D Medical Imaging Segmentation Methods}: This section aims to conduct a comprehensive report and elucidation of the various techniques proposed for 3D medical imaging segmentation. It includes detailed subsections dedicated to each category of these techniques, providing in-depth analysis of their respective merits, drawbacks, and contributions to the field. Furthermore, this section offers a meticulous exposition of representative methods, offering a comprehensive understanding of their underlying principles and functionalities.
    \item \textbf{Experiments}: This section conducted multiple experiments using state-of-the-art 3D medical imaging segmentation methods on prominent public datasets to obtain quantitative results regarding the performance of each model on each dataset. Subsequently, we performed a comprehensive analysis of the obtained results.
    \item \textbf{Discussion}: This section will critically examine the potential limitations and challenges associated with existing methods employed in 3D medical imaging segmentation. It will include a comprehensive discussion, addressing various aspects including, but not limited to, the drawbacks inherent in these methods, the limitations pertaining to their application, as well as potential ethical and security concerns associated with medical imaging datasets. Through this comprehensive analysis, we aim to shed light on the areas that necessitate further exploration and improvement within the field.
    \item \textbf{Future Works}: This section will engage in a comprehensive discussion on the potential future directions of this field, with a particular emphasis on the contributions and solutions offered by our proposed novel pipeline. It will address the limitations identified in previous methods and highlight how our approach can effectively address these challenges. Additionally, it will explore the potential of incorporating generative models, such as the diffusion model, to tackle the long-tail problem by generating mask-image pair for minority classes. Furthermore, it will investigate the potential benefits of large-scale self-supervised pre-training for feature extraction, enhancing the overall performance of the segmentation task. By delving into these promising avenues, we aim to pave the way for further advancements and improvements in 3D medical imaging segmentation.
    \item \textbf{Conclusion}: This section will provide a comprehensive conclusion on the current existing methods for volumetric medical imaging segmentation, highlighting their contributions and benefits. It will synthesize the findings and insights gained from the previous sections to assess the overall progress made in this field. Additionally, it will emphasize the unique contribution and benefits offered by our proposed method, showcasing its potential to overcome the limitations and enhance the accuracy and efficiency of volumetric medical imaging segmentation. Through this conclusive analysis, we aim to provide a clear and comprehensive understanding of the existing methods and position our proposed approach as a promising solution for future research and practical applications in this domain.
    
\end{itemize}

\section{Background}

\subsection{Mechanism of NCCT}

\begin{figure}
  \centering

  \begin{subfigure}{0.3\linewidth}
    \centering
    \includegraphics[width=\linewidth]{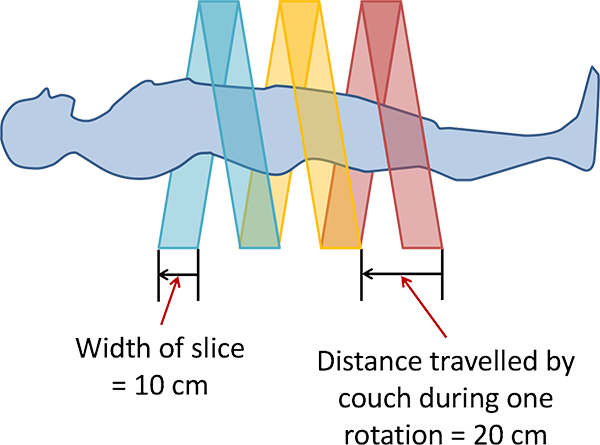}
    \caption{}
    \label{fig:a}
  \end{subfigure}
  \hfill
  \begin{subfigure}{0.3\linewidth}
    \centering
    \includegraphics[width=\linewidth]{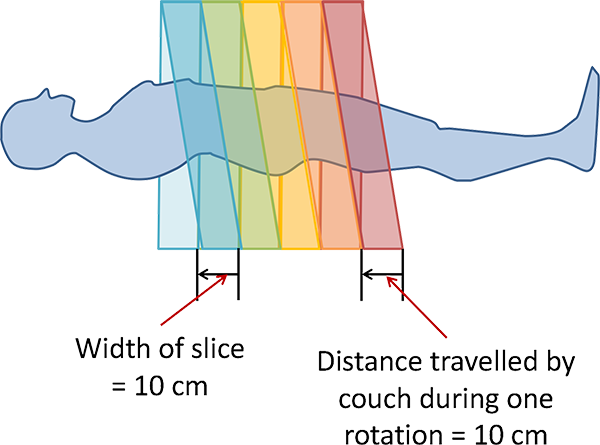}
    \caption{}
    \label{fig:b}
  \end{subfigure}
  \hfill
  \begin{subfigure}{0.3\linewidth}
    \centering
    \includegraphics[width=\linewidth]{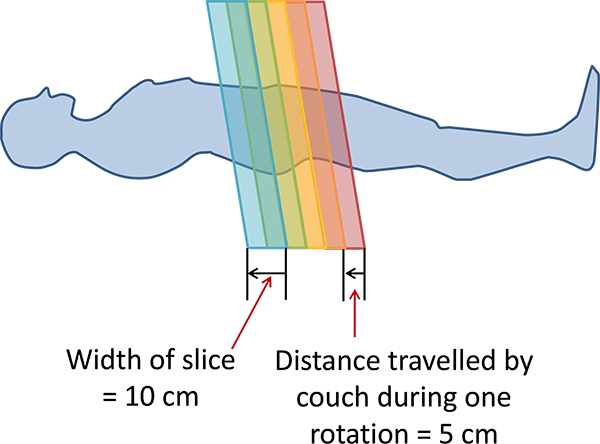}
    \caption{}
    \label{fig:c}
  \end{subfigure}

  \caption{Pitch parameter in NCCT. In \ref{fig:a}: Pitch = 20/10 = 2. In \ref{fig:b}: Pitch = 10/10 = 1. In \ref{fig:c}: Pitch = 5/10 = 0.5.}
  \label{fig:2}
\end{figure}

During the CT scanning process, the patient is positioned on the CT scanner table and gradually moved through the gantry, which houses the X-ray tube and detector array. The X-ray tube rotates around the patient, emitting X-rays that traverse through the body and are captured by the detectors positioned on the opposite side. This rotation and movement of the table are controlled by the pitch parameter \cite{39} (shown in equation \ref{eq:1}), which determines the distance the CT scanner table travels during each rotation of the X-ray tube, shown in Figure \ref{fig:2}. If the pitch number is greater than 1, it means that the couch travels more than the width of the beam, i.e. there are gaps. In contrast, if the pitch number is less than 1, it means that the couch travels less than the width of the beam, i.e. there is overlap.

\begin{equation}
Pitch = Distance \hspace{0.1cm} couch \hspace{0.1cm} travels \hspace{0.1cm} / \hspace{0.1cm} Width \hspace{0.1cm} of \hspace{0.1cm} slice
\label{eq:1}
\end{equation}

\begin{figure}
    \centering
    \includegraphics[width=0.9\linewidth]{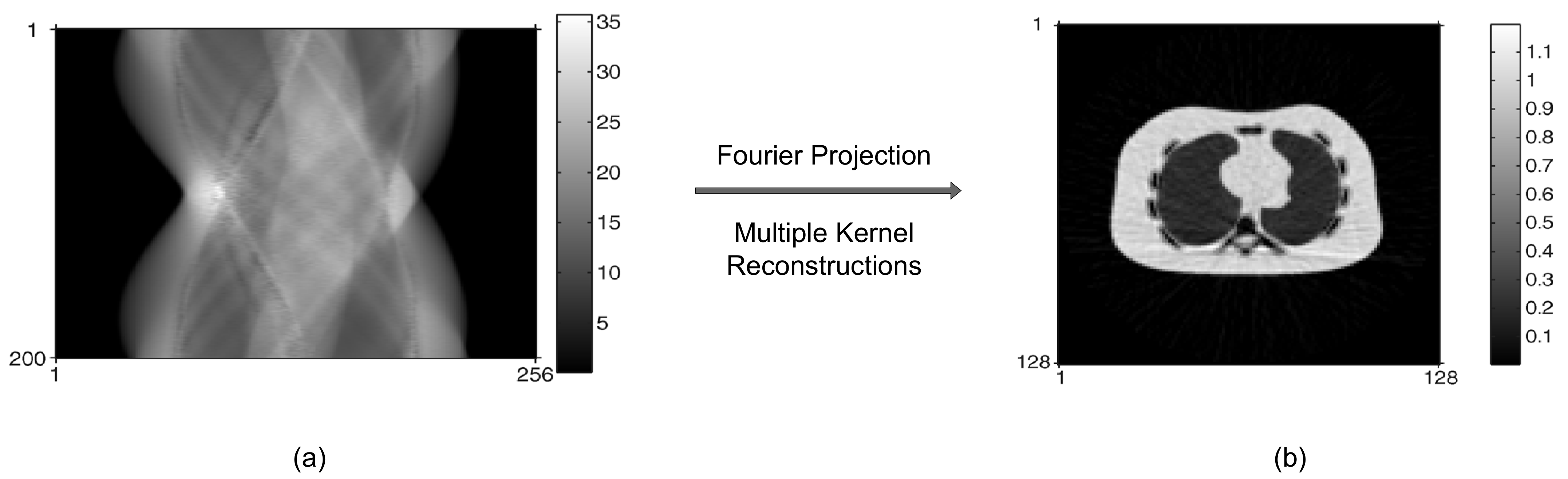} 
    \caption{Process of Prospective Reconstruction. Figure (a) shows a sinogram and figure (b) shows its axial view of reconstructed CT scan. \cite{41}}
    \label{fig:3}
\end{figure}

As the X-ray beam passes through the body, it interacts with different tissues, resulting in varying levels of X-ray attenuation. The detectors measure the X-ray attenuation at each angle, generating a series of raw data projections known as sinograms \cite{40}, shown in sub-figure (a) of Figure \ref{fig:3}.

After obtaining the raw data sinograms, the next step involves prospective reconstruction \cite{42}, shown in Figure \ref{fig:3}. This process entails performing a Fourier projection on the sinogram and applying specific kernels tailored to different types of tissues to the projected image. The primary objective is to optimize image quality and enhance the contrast between various structures, such as bones, soft tissues, and more. This technique is commonly referred to as multiple kernel reconstructions \cite{43}. During the prospective reconstruction, the pitch value plays a crucial role in determining the thickness of the resulting CT slices. A smaller pitch value indicates a smaller distance traveled by the table per rotation, resulting in thinner CT slices. Conversely, a larger pitch value leads to thicker CT slices. This parameter directly influences the spatial resolution and image quality of the CT scans. The thin slices, in another words, is obtained from the prospective reconstruction of thin scans.

\begin{figure}
    \centering
    \includegraphics[width=0.4\linewidth]{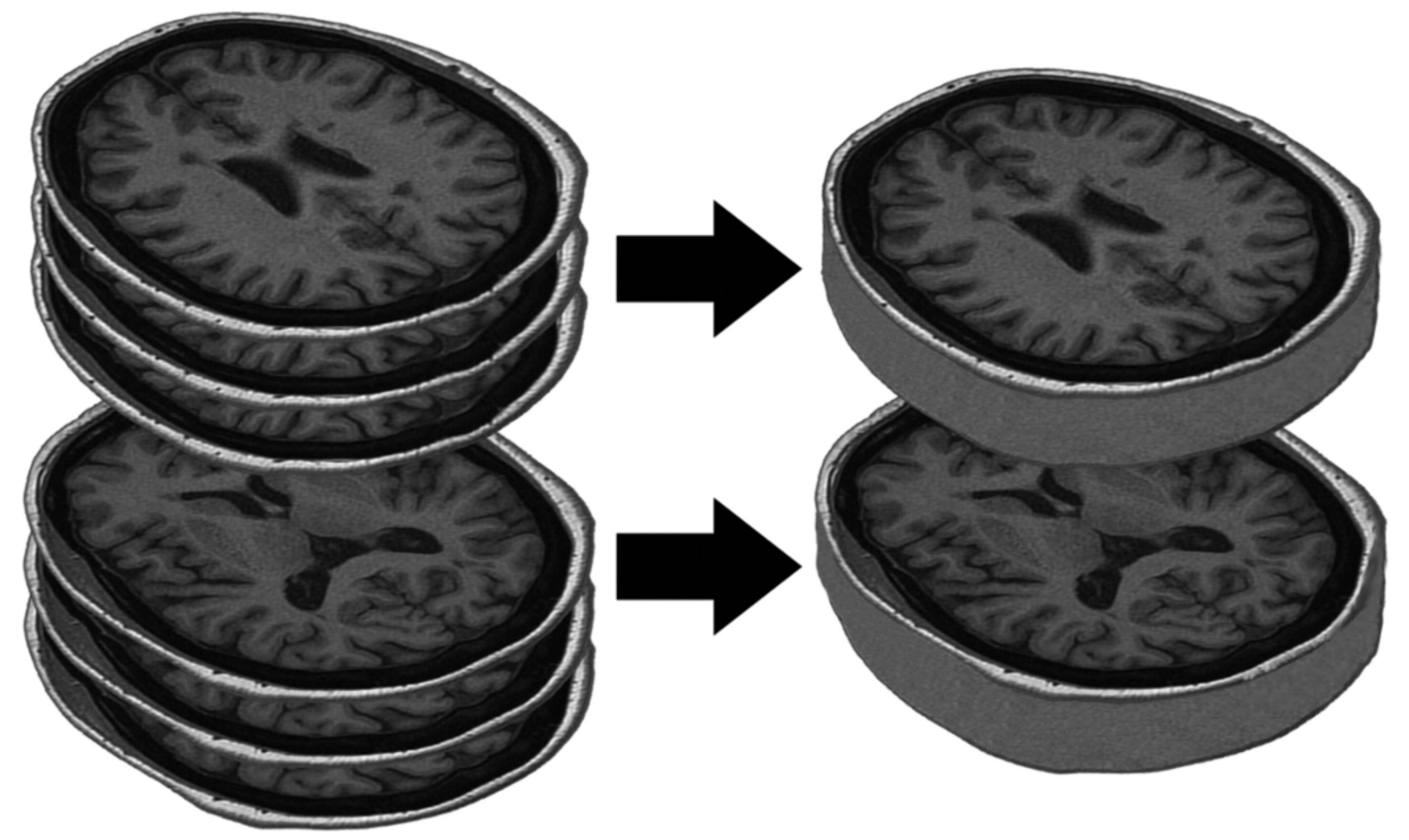} 
    \caption{Averaging of three sequential slices. \cite{45}}
    \label{fig:4}
\end{figure}

\begin{figure}
    \centering
    \includegraphics[width=0.6\linewidth]{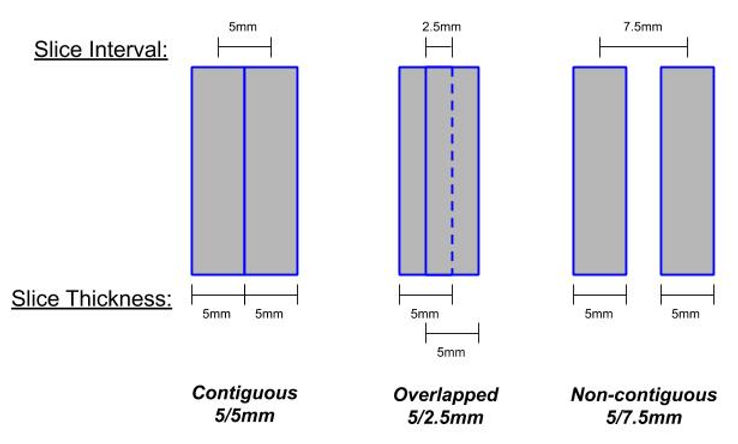} 
    \caption{Different interval strategies for averaging thin slices}
    \label{fig:5}
\end{figure}

After acquiring thin slices, a retrospective reconstruction \cite{37} technique can be employed to obtain thick slices. This process involves utilizing an averaging algorithm, such as the average intensity projection (AIP) \cite{44}, to merge multiple thin slices into a single thick slice, shown in Figure \ref{fig:4}. The thick slices can be generated with varying thicknesses and from different orientations, including axial, sagittal, and coronal planes \cite{46}. Additionally, there are three different interval strategies for averaging thin slices into thick slices: contiguous, non-contiguous, and overlapped intervals, shown in Figure \ref{fig:5}.

\subsection{Image Segmentation in Computer Vision}

Image segmentation \cite{7} is a foundational and extensively investigated problem in the field of computer vision (CV). Its objective is to partition an image into distinct and non-overlapping regions, with each pixel assigned a corresponding label. This task entails a pixel-level classification challenge, which inherently introduces a heightened level of complexity when compared to other CV problems, including image-level classification or object detection. Image segmentation can be classify into different categories according to tasks, which includes foreground-background segmentation, semantic segmentation, instance segmentation, and panoptic segmentation.

\begin{itemize}
    \item \textbf{Foreground-background segmentation} aims to separate the main subject (foreground) from the surrounding background in an image.
    \item \textbf{Semantic segmentation} Involves assigning semantic labels to each pixel in an image, enabling the identification and differentiation of various object classes and regions.
    \item \textbf{Instance segmentation} is the process of distinguishing individual object instances. This can involve either disregarding the semantic labels of each instance and segmenting entire images or assigning semantic labels to the instances while excluding the background.

    \item \textbf{Panoptic segmentation} combines the concepts of semantic and instance segmentation to provide a unified framework for labeling both object classes and individual instances, encompassing the entire visual scene.
\end{itemize}

In the domain of medical imaging segmentation, the prevalent tasks primarily revolve around foreground-background segmentation or semantic segmentation, with the choice between them being determined by the specific task at hand and the characteristics of the dataset being used.

\subsection{Dataset and Benchmark}

The availability of pixel-level annotated NCCT datasets is relatively limited compared to datasets for classification and detection tasks \cite{zhang2024meddet,cai2024medical,cai2024msdet}, primarily due to the substantial cost and time investment required for radiologists to annotate three-dimensional volumes. Nevertheless, the advent of machine learning in medical imaging analysis has led to an increase in the number of segmentation datasets. These datasets are predominantly sourced from challenges such as Grand Challenge, MICCAI Challenge, and ISBI Challenge, while others may be curated by organizations such as RSNA or SIIM-ACR. Table \ref{table:1} presents a compilation of prominent and widely utilized datasets and benchmarks in the domain of 3D medical imaging segmentation.

\begin{table*}
\centering 
\begin{tabular}{p{0.2\textwidth}p{0.2\textwidth}p{0.2\textwidth}p{0.2\textwidth}} 
\hline\hline 
Name & Label & Modality & Content \\ 
\hline 
MSD \cite{47} & Semantic & NCCT, MRI & Multi-organ \\
BCV & Semantic & NCCT & Multi-organ \\
IHM \cite{48} & fore-background & NCCT & Brain Hemorrhage \\
INSTANCE \cite{49} & fore-background & NCCT & Brain Hemorrhage \\
RICORD \cite{50} & fore-background & NCCT & Chest \\
HECKTOR \cite{51} & fore-background & NCCT & Neck Tumor \\
AMOS \cite{52} & Semantic & NCCT, MRI & Multi-organ\\

\hline 
\end{tabular}
\caption{Description of the features of Public Dataset} 
\label{table:1} 
\end{table*}

\section{3D Medical Imaging Segmentation Methods}

In this section, a comprehensive report and analysis of the current state-of-the-art techniques proposed for 3D medical imaging segmentation, with a specific focus on NCCT, is conducted. The methods under investigation are categorized into two distinct categories based on their architectures: Convolutional Methods and Convolutional-Transformer Hybrid Methods, shown in Table \ref{table:2}. The report encompasses a detailed explanation and critical evaluation of each category, including their underlying principles, advantages, limitations, and notable contributions to the field. The objective is to provide readers with a thorough understanding of the existing approaches and their implications for 3D medical imaging segmentation, with particular emphasis on NCCT.

\begin{table*}
\centering 
\begin{tabular}{p{0.15\textwidth}p{0.3\textwidth}p{0.2\textwidth}} 
\hline\hline 
Model & Architecture & Dimension \\ 
\hline 
3D U-Net \cite{26} & Convolutional & 3D \\
nnUNet \cite{27} & Convolutional & 3D \\
TransUNet \cite{34} & Convolutional+Transformer & 2D Reconstruction \\
UNETR \cite{53} & Convolutional+Transformer & 3D \\
nnFormer \cite{35} & Convolutional+Transformer & 3D \\

\hline 
\end{tabular}
\caption{Representative methods for 3D medical imaging segmentation.} 
\label{table:2} 
\end{table*}

\subsection{Convolutional Methods}

With the notable success of fully convolutional networks (FCNs) \cite{54} in the domain of semantic segmentation for general vision tasks, Ronneberger et al. introduced a specialized variant called U-Net \cite{24} specifically designed for biomedical imaging segmentation. Unlike previous convolutional neural network architectures proposed for medical imaging segmentation, such as the one presented by Ciresan et al. \cite{55}, U-Net addresses several limitations. Firstly, the previous network approach suffers from computational inefficiency as it requires separate execution for each patch, resulting in redundant computations due to overlapping patches. Secondly, there exists a trade-off between localization accuracy and contextual information utilization. Larger patches necessitate the inclusion of more max-pooling layers, which compromises localization accuracy, whereas smaller patches limit contextual awareness. In contrast, the U-Net architecture offers remarkable simplicity and ease of implementation. Additionally, it demonstrates enhanced efficiency, with segmentation of a 512x512 image requiring less than a second on a modern GPU. Consequently, U-Net has emerged as one of the most widely adopted backbone architectures for medical imaging segmentation, maintaining its relevance over an extended period of time \cite{56, 57}.

\begin{figure}
  \centering

  \begin{subfigure}{0.4\linewidth}
    \centering
    \includegraphics[width=\linewidth]{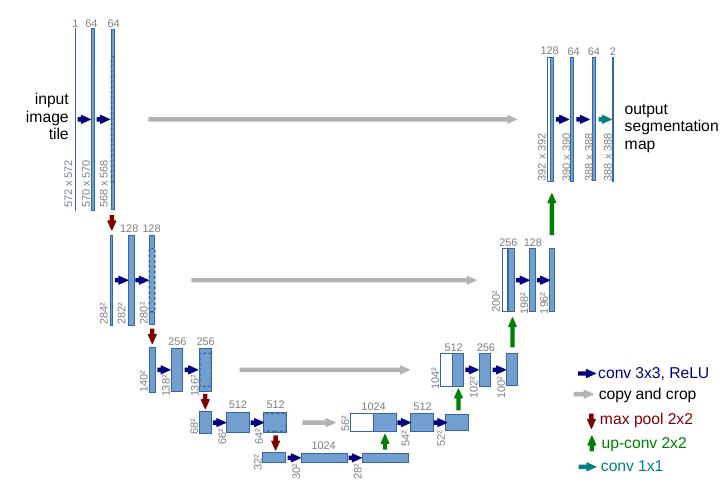}
    \caption{}
    \label{fig:aa}
  \end{subfigure}
  \hfill
  \begin{subfigure}{0.4\linewidth}
    \centering
    \includegraphics[width=\linewidth]{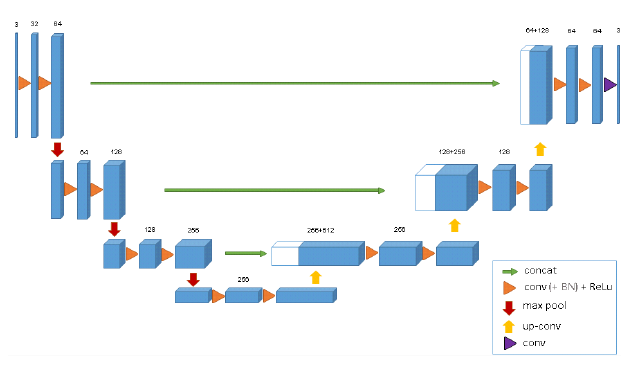}
    \caption{}
    \label{fig:bb}
  \end{subfigure}

  \caption{Comparison of U-Net and 3D U-Net architecture. In \ref{fig:aa} shows U-Net. In \ref{fig:bb} shows 3D U-Net.}
  \label{fig:6}
\end{figure}

Motivated by the remarkable success of U-Net in 2D medical imaging tasks, Cicek et al. introduced the 3D U-Net architecture \cite{26}, specifically tailored for 3D biomedical imaging segmentation. The 3D U-Net architecture retains the simplicity and straightforwardness of the original U-Net, with the key modification involving the replacement of the 2D convolutional kernels with 3D convolutional kernels, shown in Figure \ref{fig:6}. Similar to the vanilla U-Net, the 3D U-Net comprises a total of 9 stages, including 5 stages in the encoder and 4 stages in the decoder. Each encoder stage consists of the input followed by two consecutive 3x3x3 convolutional kernels and a 2x2x2 max pooling operation. Similarly, each decoder stage consists of the input followed by two consecutive 3x3x3 convolutional kernels, but with the addition of a 2x nearest neighbor interpolation for upsampling. This straightforward adaptation allows the 3D U-Net to leverage the volumetric information inherent in 3D medical images and has shown promising results in various biomedical imaging segmentation tasks. The original contribution of the 3D U-Net was focused on enabling dense predictions using sparse pixel-level annotations. Subsequently, it has been demonstrated that the 3D U-Net architecture is also well-suited for 3D medical imaging segmentation tasks, including the segmentation of NCCT images \cite{58, 59}. 
The 3D U-Net inherits the advantageous characteristics of the vanilla U-Net, such as its simplicity, efficiency, and speed. However, it also highlights a significant limitation of the vanilla 3D U-Net: the absence of a comprehensive pipeline for medical imaging segmentation and a disregard for the importance of data preprocessing and augmentation. Consequently, more contemporary alternative models have emerged to address these shortcomings and offer improved performance in medical imaging segmentation tasks.

\begin{figure}
    \centering
    \includegraphics[width=0.9\linewidth]{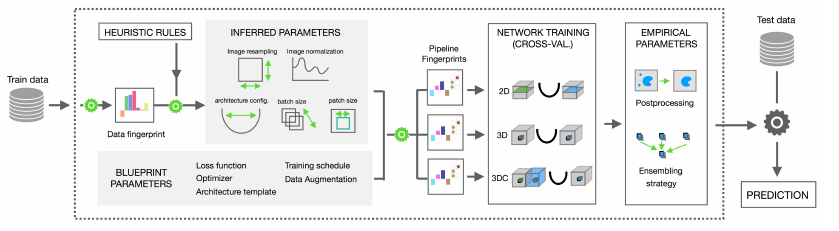} 
    \caption{Novel pipeline of nnUNet.}
    \label{fig:7}
\end{figure}

Inspired by the vanilla 3D U-Net, Isensee et al. introduced nnUNet \cite{27}, which utilizes the 3D U-Net as a backbone while placing greater emphasis on data preprocessing, data augmentation, and hyperparameter tuning of the pipeline, which shown in Figure \ref{fig:7}. This approach has proven to be effective in refining the original backbone and achieving significant improvements over the vanilla 3D U-Net. The modifications and contributions of nnUNet can be summarized as follows: Firstly, nnUNet incorporates a unique heuristic data preprocessing technique that involves resampling and normalizing the input 3D volume data, as well as considering the affine metadata, which provides spatial information regarding the pixel-to-real-world scale in medical imaging. This preprocessing approach enables the model to better capture the spatial characteristics of the 3D volume, leading to enhanced feature extraction. Secondly, the network incorporates meta-learning techniques, specifically Network Architecture Search (NAS), which automates the tuning of critical hyperparameters such as batch size, patch size, learning rate, and optimizer. This automated hyperparameter optimization not only saves time compared to manual grid search but also increases the likelihood of finding an optimal set of hyperparameters for the model. Furthermore, nnUNet exhibits versatility by being applicable not only to NCCT images but also to contrast-enhanced computed tomography (CECT) and MRI. The innovative data preprocessing approach employed in nnUNet is considered state-of-the-art in 3D medical imaging and has been widely adopted in subsequent models for 3D medical imaging segmentation \cite{35}. However, one potential limitation of nnUNet is the use of a relatively older backbone architecture, 3D U-Net, which hinders its feature extraction capabilities and segmentation performance compared to more recent developments in transformer and self-attention architectures \cite{34,53,35}.

\subsection{Convolutional-Transformer Hybrid Methods}

\subsubsection{2D Reconstruction Method}

The recent integration of transformer architecture and self-attention mechanisms into the field of computer vision \cite{28} has sparked significant interest in their application to medical imaging segmentation. Chen et al. introduced TransUNet \cite{34}, which is the first medical image segmentation framework that incorporates self-attention mechanisms in a sequence-to-sequence prediction framework. Recognizing the loss of feature resolution introduced by Transformers, TransUNet adopts a hybrid CNN-Transformer architecture that combines detailed high-resolution spatial information from CNN features with the global context encoded by Transformers. In this approach, a self-attention layer is appended at the end of the pyramid CNN decoder. Leveraging the principles of the 3D U-Net \cite{26}, the self-attentive features encoded by Transformers are upsampled and combined with various high-resolution CNN features derived from the encoding path, facilitating precise localization in the segmentation task. 
The main contribution of TransUNet is the incorporation of a self-attention mechanism in the encoder of a medical segmentation model. However, a limitation of this approach is that the self-attention layer only attends to the last layer of the encoder output, which restricts its ability to capture a comprehensive range of image features. This constraint diminishes the potential advantages offered by the transformer architecture, thereby limiting the performance of the encoder and the effectiveness of feature extraction. Moreover, TransUNet is essentially a 2D segmentation model that reconstructs 3D segmentation slice-by-slice. This introduces a significant drawback as the model cannot learn the continuity between adjacent slices and fails to capture the spatial information inherent in a 3D volume. Consequently, the resulting segmentation mask may exhibit irregularities and appear fragmented when viewed from sagittal and coronal perspectives.

\begin{figure}
    \centering
    \includegraphics[width=0.7\linewidth]{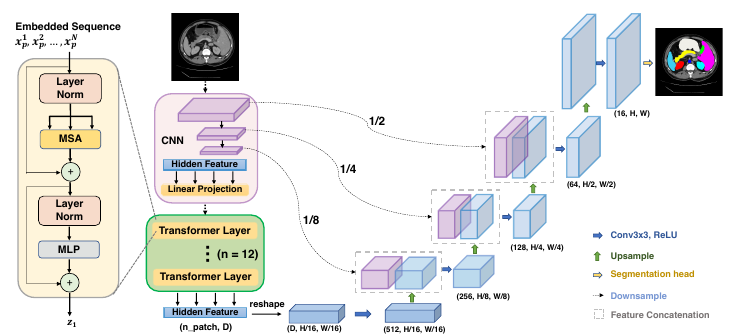} 
    \caption{The hybrid CNN-Transformer architecture of TransUNet, which demonstrates the self-attention layers only attend the output of CNN encoder.}
    \label{fig:8}
\end{figure}

\subsubsection{3D Methods}

\begin{figure}
    \centering
    \includegraphics[width=0.7\linewidth]{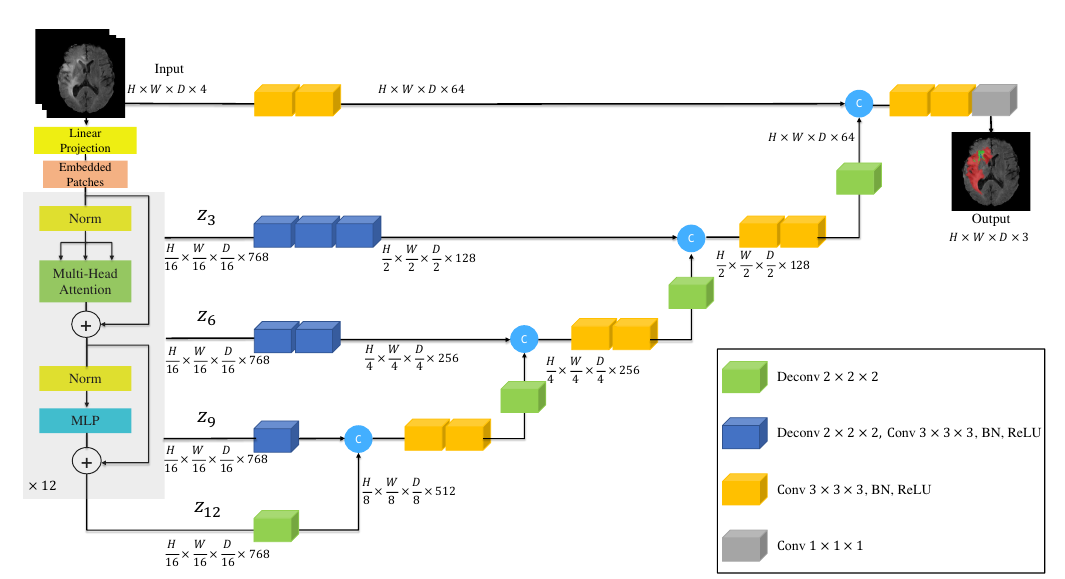} 
    \caption{The transformer encoder and convolutional decoder of UNETR.}
    \label{fig:9}
\end{figure}

The 3D U-Net \cite{26} has demonstrated its effectiveness as a fully convolutional neural network (FCN) \cite{54} for representation learning. However, FCN-based approaches are limited in their ability to capture long-range dependencies due to their localized receptive field. In contrast, Transformer-based computer vision models have successfully addressed this limitation. Building upon the FCN and the vanilla Vision Transformer (ViT) \cite{28}, Hatamizadeh et al. proposed a novel 3D medical imaging segmentation model, named UNETR \cite{53}. The model consists of a Transformer encoder and a 3D convolutional decoder, shown in Figure \ref{fig:9}. Similar to ViT, the 3D volume is divided into non-overlapping 3D patches, which are then flattened into a 1D vector and processed by a patch embedding layer. Additionally, a 1D learnable positional embedding is incorporated into the embedded vector. The embedded vector is then passed through a stack of Transformer layers in the encoder. At the bottleneck of the encoder, a deconvolutional layer is applied for upsampling the transformed feature map, effectively increasing its resolution by a factor of 2. The resized feature map is concatenated with the feature map from the previous transformer output and fed into consecutive 3×3×3 convolutional layers in the decoder. Finally, the output is upsampled using a deconvolutional layer. The primary contribution of this model lies in its exclusive utilization of Transformers as an Encoder, effectively harnessing their ability to capture long-range dependencies. However, an evident drawback of this model is its continued reliance on a convolutional decoder, which introduces relatively high complexity compared to a lightweight MLP or attention decoder. Consequently, this increases the training time required for the model.

\begin{figure}
    \centering
    \includegraphics[width=0.7\linewidth]{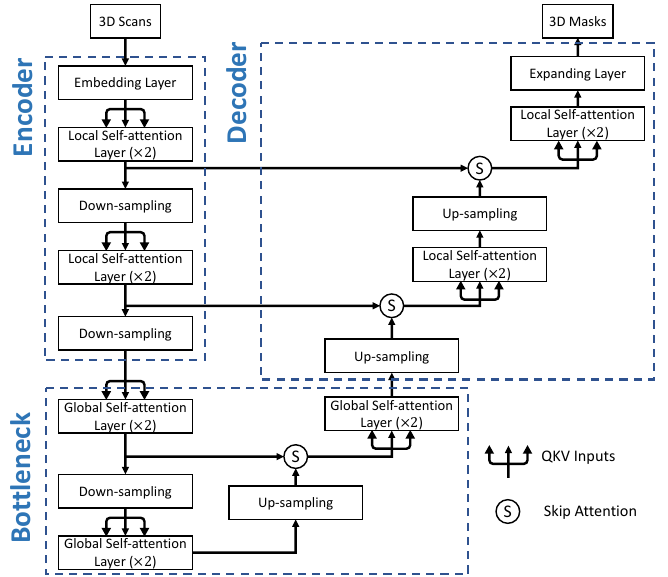} 
    \caption{The architecture of nnFormer.}
    \label{fig:10}
\end{figure}

Zhou et al. proposed nnFormer, a CNN-Transformer hybrid model for 3D medical imaging segmentation. In contrast to vanilla ViT \cite{28} and Swin Transformer \cite{31}, which utilize large convolutional kernels in the embedding block for feature extraction, nnFormer discovered that employing successive convolutional layers with small kernel sizes at the embedding layer of the encoder yields more advantages in the initial stage. This approach contributes in two significant ways: Firstly, the successive convolutional embedding layers enable more precise encoding of pixel-level spatial information compared to patch-wise positional encoding used in vanilla ViT. Secondly, the use of small kernel sizes in the embedding layers helps reduce computational complexity while maintaining equal-sized receptive fields. Moreover, nnFormer incorporates two successive transformer layers in each block of the encoder. The first transformer layer consists of a Local Volume-based Multi-head Self-attention (LV-MSA) followed by an MLP layer, while the second transformer layer comprises a Shifted Local Volume-based Multi-head Self-attention (SLV-MSA) followed by an MLP. In the decoder stage, nnFormer also employs a Transformer architecture. The structure of the two transformer blocks in the decoder closely symmetrical to those in the encoder. The architecture of nnFormer is shown in Figure \ref{fig:10} However, a limitation of this architecture design is its continued heavy reliance on the feature extraction capabilities of the embedding layer, rather than fully exploiting the potential of the Transformer itself.

\section{Experiments}

The lack of homogeneity in the comparative experiments conducted poses a significant challenge in accurately assessing the performance of different state-of-the-art methods. Consequently, directly comparing the results obtained by each method becomes infeasible. Hence, the primary objective of this experimental section is to provide the reader with a comparison under unified, fair, and equal conditions for all methods. This comparison aims to offer a convenient and accessible means of understanding the current state-of-the-art methods in the field and their relative quality compared to others. To achieve this goal, we conducted a series of experiments following specific guidelines aimed at addressing the aforementioned comparison issues. This selection of experiments for evaluation was designed to be representative of all the categories introduced in our methodology.

\subsection{Datasets}

The evaluation of the models and demonstration of their benefits are conducted using two widely recognized multi-organ segmentation datasets: Multi-Atlas Labeling Beyond the Cranial Vault (BCV) and Abdominal CT Organ Segmentation (AMOS) \cite{52}. The BCV dataset consists of 50 abdomen CT scans encompassing 13 distinct organs, while the AMOS dataset comprises 500 CT scans and 100 MRI scans involving 15 distinct abdominal organs. By utilizing these datasets as benchmarks, we aim to comprehensively evaluate the performance of each model and showcase their advantages.

\subsection{Validation Strategy}

For each model, we conducted training for 1000 epochs with a batch size of 250. The standard validation strategy involved employing a hold-out method, wherein the dataset was divided into a training set and a validation set following the guidelines provided by the respective dataset. The optimizer used for training all networks, except nnUNet, was AdamW \cite{61}. During both training and inference, the data was resampled to achieve an isotropic spacing of 1.0 mm (with results reported on the original spacing). The input patch size for 3D networks was set to 128×128×128, while for 2D networks, it was set to 512×512. The corresponding batch sizes were 2 and 14 for 3D and 2D networks, respectively.

\subsection{Evaluation Metrices}

In the domain of 3D semantic segmentation in medical imaging, the evaluation of segmentation performance relies on two widely recognized metrics: the Dice Similarity Coefficient (DSC) \cite{62} and the Surface Dice Similarity (SDC). The DSC calculates the intersection over union between the predicted and ground-truth segmentation masks, providing a measure of their similarity. On the other hand, the SDC quantifies the similarity between the predicted and ground-truth masks, specifically focusing on the surface regions. These evaluation metrics are widely adopted in the medical imaging field to assess the accuracy and quality of segmentation outcomes. In our proposed model evaluation, we will employ these metrics to evaluate the performance of our approach.

\subsubsection{Dice Similarity Coefficient (DSC)}

The Dice Similarity Coefficient (DSC) is a commonly utilized evaluation metric in medical imaging segmentation applications. It quantifies the degree of overlap between the predicted segmentation and the corresponding ground truth segmentation. The DSC value ranges between 0 and 1, with 0 indicating no overlap and 1 indicating a perfect match between the predicted and ground truth segmentations.

\begin{equation}
  DSC = (2 \times Intersection)/Union
  \label{eq:2}
\end{equation}

The calculation of DSC involves determining the intersection of the predicted and ground truth segmentations and dividing it by the union of their respective areas. The $Intersection$ represents the common region between the two segmentations, while the $Union$ represents the combined area of both the predicted and ground truth segmentations.

The DSC is particularly valuable in assessing the accuracy of segmentation models, especially in situations where there is a substantial class imbalance. It provides a comprehensive measure of the segmentation performance, accounting for both the presence and absence of the segmented regions.

\subsubsection{Surface Dice Similarity (SDC)}

The SDC takes into account the precision of the surface boundaries between the predicted and ground truth segmentations. It evaluates the agreement in terms of the boundaries, which are of particular importance in medical imaging applications. The SDC provides a more localized and detailed assessment of the segmentation accuracy, capturing the intricacies of the surface boundaries.

\subsection{Quantitative Results}

The quantitative results showed in Table \ref{table:3}, showcasing the superiority of nnUNet over other methods in the multi-organ segmentation datasets. This achievement can be attributed to the effective image preprocessing techniques employed and the utilization of network architecture search based on meta learning. As a result, nnUNet maintains its position as the state-of-the-art (SOTA) method in these datasets.

\begin{table}
    \centering
    \begin{tabular}{ccccc}
        \toprule
        \hline
        \multicolumn{1}{c}{} & \multicolumn{2}{c}{BCV} & \multicolumn{2}{c}{AMOS} \\
        \cmidrule(r){2-3} \cmidrule(r){4-5}
        Models & DSC & SDC & DSC & SDC \\
        \midrule
        nnUNet & \textbf{83.56} & \textbf{86.07} & \textbf{88.88} & \textbf{91.70} \\
        UNETR & 75.06 & 75.00 & 81.98 & 82.65 \\
        TransUNet & 76.72 & 76.64 & 85.05 & 86.52 \\
        nnFormer & 80.76 & 82.73 & 84.20 & 86.38 \\
        \hline
        \bottomrule
    \end{tabular}
    \caption{Comparison of quantitative results of different models on benchmarks}
    \label{table:3}
\end{table}

\section{Discussion}

\subsection{Potential of convolutional-free architecture}

The significance of having a convolutional-free method in our encoder-decoder model cannot be overstated. In the current landscape of segmentation techniques, most methods rely heavily on convolutional-based architectures or employ CNN-Transformer hybrid models. However, our proposed approach breaks away from this trend by adopting a convolutional-free transformer architecture.

By comparing our convolutional-free transformer architecture with other techniques, we can observe several distinct advantages. Firstly, our architecture excels in capturing global context information, surpassing the capabilities of CNN-based models. The self-attention mechanism inherent in the transformer architecture enables the model to focus on important regions within the input image. This focused attention leads to more precise semantic labeling and the generation of high-quality masks.

Moreover, transformers possess the unique ability to capture long-range dependencies within the image. This feature allows for a more comprehensive modeling of complex relationships between pixels. By considering distant pixels, the transformer architecture can effectively understand the global structure and semantic coherence of the image, enhancing the segmentation performance.

The convolutional-free nature of our model not only demonstrates its capability to outperform CNN-based counterparts but also highlights the potential of transformer architectures in the field of image segmentation. The absence of convolutions in our method liberates the model from the limitations associated with local receptive fields, allowing for the exploitation of global contextual information. This innovation opens up new avenues for advancing the accuracy and effectiveness of semantic labeling tasks in medical imaging and other domains.

\subsection{Thin-thick slices domain adaptation}

The integration of domain adaptation between thin slices and thick annotations holds immense importance in the context of our proposed method. Presently, the limitation lies in the fact that all public datasets primarily consist of thick slices, rendering the existing methods applicable only to such datasets.

However, our method breaks free from this constraint by introducing a domain adaptation technique that enables the segmentation of thin slices based on thin annotations. This novel capability allows us to effectively segment thin slices with a high z-resolution, generating an isomorphic mask that faithfully represents the volume of focus. 

The implications of this advancement are significant, particularly in the fields of surgical planning and navigation. The precise segmentation of thin slices and the subsequent generation of an accurate isomorphic mask enable a more refined presurgical simulation. Surgeons can benefit from this enhanced preoperative planning by gaining a comprehensive understanding of the anatomical structures and their spatial relationships within the volume of interest.

During surgery, the isomorphic mask serves as a valuable tool, providing direct guidance and assisting in intraoperative decision-making. By accurately representing the volume of focus, our method has the potential to improve patient outcomes and reduce surgical complications. Surgeons can navigate the surgical site with greater confidence and make informed decisions, ensuring a more precise and successful surgical intervention.

Overall, the incorporation of domain adaptation between thin slices and thick annotations in our method not only expands the scope of application to thin slice datasets but also unlocks the potential for improved surgical planning, navigation, and patient outcomes.

\subsection{Ethic problem of medical imaging data}

In the domain of medical imaging, a notable concern revolves around determining the permissible sharing of data without compromising patients' privacy. Establishing clear guidelines and policies regarding the types of medical imaging data that can be shared publicly or among researchers, and defining the specific conditions under which sharing is permissible, becomes imperative. Furthermore, given the substantial size of medical imaging datasets, there exists a potential risk of data breaches and hacking incidents, which may result in the unauthorized disclosure of sensitive patient information.

To address these significant concerns, researchers must adopt additional measures to safeguard the security and privacy of medical imaging data. This entails employing secure storage systems that meet stringent security standards, implementing robust data access controls to limit unauthorized access, and ensuring the anonymization of patient data to minimize the risk of identification. Furthermore, obtaining informed consent from patients becomes an essential prerequisite before collecting and utilizing their medical data for research purposes. By placing a paramount emphasis on the ethical and security considerations surrounding medical imaging data acquisition, researchers can uphold a responsible and transparent approach in utilizing patient data.

In summary, it is crucial to establish clear guidelines and policies regarding the sharing of medical imaging data to protect patients' privacy. Researchers should prioritize the adoption of stringent security measures, including secure storage systems, data access controls, and anonymization techniques. Obtaining informed consent from patients is essential to ensure the ethical use of their medical data. By upholding these principles, researchers can ensure the responsible and transparent utilization of medical imaging data while safeguarding patient privacy.

\section{Future Works}

\subsection{Potential of using generative model to solve long-tail problem}

One potential area for future improvement in 3D multi-semantic medical imaging segmentation \cite{zhang2025gamed,tan2024segkan,tan2024segstitch} revolves around addressing the long-tail problem \cite{63,64}. This problem arises when dealing with highly imbalanced data in machine learning classification tasks, particularly in the context of multi-label and multi-class classification. Ideally, a training dataset for a deep learning model should follow a universal distribution rather than a normal distribution.

In scenarios where the data is mildly unbalanced, random downsampling \cite{65} of the majority class can be employed to achieve a better balance. However, for cases where the imbalance between the minority and majority classes is significant, downsampling may not be sufficient. In such instances, a weighted loss \cite{66} can be formulated to assign higher weight to the minority class during backpropagation. Despite these approaches, effectively addressing long-tail problems remains a challenging task.

Nevertheless, recent advancements in the field, such as the development of DiffuMask \cite{67}, a diffusion model for generating mask-image pairs, offer promising solutions for tackling long-tail classes. The utilization of DiffuMask holds potential in the field of medical imaging segmentation, particularly for pre-training purposes. By incorporating DiffuMask or similar techniques, it may be possible to generate long-tail classes and enhance the performance of models in addressing imbalanced data distributions.

In summary, the long-tail problem presents a notable challenge in 3D multi-semantic medical imaging segmentation. While existing methods such as random downsampling and weighted loss provide some mitigation strategies, there is still room for improvement. The recent introduction of DiffuMask offers a promising avenue for generating long-tail classes and can be explored in the context of medical imaging segmentation for pre-training purposes. By leveraging these advancements, researchers can strive to enhance the performance and address the imbalanced data distributions in this domain.

\subsection{Pre-train for medical imaging analysis}

One area for future exploration in 3D multi-semantic medical imaging segmentation relates to the insufficiency of pre-trained models specifically tailored to this domain, in contrast to the advancements seen in general computer vision. Despite the availability of extensive collections of unlabeled images in public datasets, the lack of suitable pre-trained models presents a significant challenge. To address this limitation, there is considerable potential in the development of large-scale self-supervised \cite{wu2024xlip} or contrastive pre-training models, building upon existing techniques like MoCo \cite{68} or MAE \cite{69}. By leveraging these techniques, it becomes possible to obtain a pre-trained encoder that can be effectively utilized as a frozen feature extractor in medical image segmentation tasks. 

By focusing on the development of domain-specific pre-training models through self-supervised or contrastive approaches, researchers can bridge the gap between the general computer vision domain and the unique requirements of 3D multi-semantic medical imaging segmentation. This strategy allows for the extraction of meaningful and relevant features from medical images, leading to improved segmentation performance. Furthermore, the availability of domain-specific pre-trained models will facilitate the transfer of knowledge and promote advancements in medical imaging segmentation.

\section{Conclusion}

In conclusion, this technical report provided a comprehensive overview of the current state of non-contrast CT image segmentation in the context of computer vision. We began by revisiting our proposed method, highlighting its potential for accurate and efficient segmentation of non-contrast CT images. 

To provide a solid foundation, we delved into the background of non-contrast CT imaging and discussed the importance of image segmentation in the field of computer vision. Understanding the underlying principles and challenges associated with this domain is crucial for the development of effective segmentation techniques.

We then conducted a thorough report of the representative methods in the report, including both convolutional-based and CNN-Transformer hybrid approaches. Each method was evaluated based on its contribution, advantages, and limitations. Notably, nnUNet emerged as the state-of-the-art method across various segmentation tasks, demonstrating its superiority and wide applicability.

Furthermore, we explored the correlation between our proposed method and the existing approaches. By leveraging the strengths of both convolutional-based models and CNN-Transformer hybrids, our method aims to enhance the global context modeling and achieve more accurate semantic labeling and mask generation.

Looking ahead, several potential future research directions in this domain were identified. These include addressing the long-tail problem in multi-semantic segmentation, exploring the use of pre-trained models specifically designed for medical imaging, and investigating the applicability of self-supervised or contrastive pre-training techniques. These avenues hold promise for further advancements in non-contrast CT image segmentation, enabling more precise presurgical simulations, improved patient outcomes, and reduced surgical complications.

In summary, this technical report has provided a comprehensive understanding of non-contrast CT image segmentation, highlighted the strengths and weaknesses of current methods, and identified potential areas for future research. By continuing to explore and refine these techniques, we can pave the way for more accurate and efficient segmentation methods in the field of medical imaging.

\section{Acknowledgements}
I would like to acknowledge the support of Zeyu Zhang and AI Geeks in contributing to this research project.


\end{document}